\documentclass[doublecol]{epl2}
\usepackage{amssymb}
\usepackage{amsmath}
\usepackage{graphicx}

\institute{
  \inst{1} College of Physics, Communication and Electrons , Jiangxi Normal University, Nanchang, 330022, China\\
  \inst{2} Center for Engineered Quantum Systems, School of Mathematics and Physics, The University of Queensland, St Lucia, Queensland 4072, Australia
}
{Nanoelectromechanical systems (NEMS)}
{Quantum state engineering and measurements}
{Structural acoustics and vibration}
\abstract{In this study, we investigate the phonon antibunching effect in a
coupled nonlinear micro/nanoelectromechanical system (MEMS/NEMS) resonator at a finite
temperature. In the weak driving limit, the optimal condition for phonon antibunching is given by
solving the stationary Liouville-Von Neumann master equation. We show that
at low temperature, the phonon antibunching effect occurs in the regime of weak
nonlinearity and mechanical coupling, which is confirmed by analytical and numerical solutions. We also
find that thermal noise can degrade or even destroy the antibunching effect for different mechanical coupling strengths. Furthermore, a transition from strong antibunching to bunching for phonon correlation has been observed in the temperature domain. Finally, we find that a suitably strong driving in the finite-temperature case would help to preserve an optimal phonon correlation against thermal noise.}

\begin{document}

\title{Phonon antibunching effect in coupled nonlinear
micro/nanomechanical resonator at finite temperature}
\author{Shengguo Guan$^1$, Warwick P. Bowen$^2$, Cunjin Liu$^1$, and Zhenglu
Duan$^{1a}$}
\maketitle

\footnotetext[1]{$^a$ duanzhenglu@jxnu.edu.cn}

\section{Introduction}

Quantum state transfer and storage are crucial in quantum information
processing. To date, the photon has been the information carrier most commonly used to transfer and
store quantum information, and it has the advantage of high
velocity, robustness to different environments, and good integrability. However,
phonons, which are vibrational modes of mechanical resonators, can be maintained for a
very long time before being eventually damped, and they have the ability to
interact with a wide range of quantum systems, such as electric, magnetic,
and optical systems. Therefore, phonons also have promising potential as
quantum information carriers \cite{PI,PI1,PI2}.

In quantum phononic networks, the nonclassical states of phonons or of a single
phonon are important elements. Many methods to prepare nonclassical states
of phonons or a single phonon have been proposed. For example, a single-phonon
Fock state is prepared by two-phonon damping \cite{SP0}, a non-Gaussian
state of a mechanical resonator is generated by performing measurements \cite{NCP}, and a
single phonon is produced by the heralded measurement of the Stokes photon in
cavity optomechanics \cite{SP1,NCP1}.

It is well known that a sufficiently high Kerr nonlinearity in an optical
cavity will prevent further photons from entering once one photon is present, and this is called the photon blockade effect \cite{PB}. The transmitted light
passing through the optical cavity then shows strong antibunching, which can
be used to convert a coherent field into a train of single photons \cite%
{photonB}. Similar to its optical counterpart, the phonon can also exhibit an
antibunching effect in a strong nonlinear mechanical resonator \cite%
{Phonon,Phonon0,Phonon1,Phonon2,Phonon3,Phonon4}. However, the typical
intrinsic nonlinearity of most micro/nanomechanical resonators is usually
very weak\cite{N0,PC,mems,N1,N2,N3}. Thus, we aim to determine whether strong phonon antibunching can be realized in a coupled micro/nanomechanical
resonator system with weak intrinsic mechanical nonlinearity.

In fact, a similar problem of weak nonlinearity also exists in the optical
case. To address this problem, Liew and Savona found that
photons exhibit strong antibunching in a system consisting of two coupled
optical cavities with weak optical nonlinearity, which is called the
unconventional photon blockade (UPB). Ciuti \textit{et al.} found that the
underlying physics of UPB is that the excitation from the vacuum state to
the two-photon state was suppressed by the destructive quantum interference
between distinct pathways with a small and finite nonlinearity in the
auxiliary cavity \cite{UPB0,UPB,UPB1,UPB2,UPB3}.

Owing to the high resonant frequency of an optical cavity, the thermal
photon number is negligible at room temperature. Therefore, the effect
of thermal photons on the photon statistics and correlations is usually
neglected safely. However, for mechanical resonators, the thermal phonon
number is not negligible, and significantly influences the phonon
blockade even at temperatures of the order of several milliKelvins \cite{Phonon0,Phonon1,T}. Hence, an
analytical study of phonon antibunching may help to give a deeper insight
into the quantum correlation at finite temperature.

Motivated by these above-mentioned studies, we propose a scheme to realize
a strong phonon antibunching effect by coupling a linear micro/nanoelectromechanical
system (MEMS/NEMS) resonator to a weakly nonlinear MEMS/NEMS resonator.
The optimal conditions required to observe strong phonon antibunching in this
system are analytically found at finite temperature based on the stationary
Liouville-Von Neumann master equation. In particular, we systematically
study the effect of the thermal noise on the quantum statistic and
correlation properties. We expect our system to be useful for generating a
nonclassical phonon state at finite temperature.

\section{Model}

\label{model} As shown in Fig. 1, the system under consideration consists of
two linearly coupled doubly clamped  mechanical beams. One mechanical beam
(referred to as resonator $1$) is a linear resonator coherently driven by a
force signal, and the other (referred to as resonator $2$) contains a weak
Duffing nonlinearity without driving. The mechanical nonlinearity can be
intrinsic, such as geometric and material nonlinearity \cite{N0,mems}, or
it can be induced by coupling the mechanical oscillator to a low-dimensional auxiliary
system, such as a plate capacitance\cite{PC}, Cooper-pair boxes \cite{Qubit},
polar molecules \cite{PM}, and quantum dots \cite{QD}. As an example, the coupling between
mechanical resonators could be achieved by applying a voltage
between two electrodes patterned on the beams, which gives rise to a static
intermodal coupling \cite{coupling0,coupling1}, or by using a non-rigid
anchor strain to mediate between the two beams \cite{coupling2}. As opposed to
the latter case, dielectric intermode coupling can be tuned freely.
\begin{figure}[tbp]
\onefigure[width=3.2in]{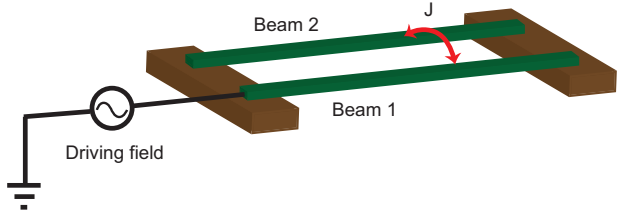}
\caption{{\ (Color online) Schematic of the phonon blockade effect.
A linear mechanical beam that is driven by an external force is linearly coupled
to another nonlinear mechanical beam.}}
\label{fig1}
\end{figure}

We assume that the resonator $1$ is harmonically driven by an external force
with amplitude $F$ and frequency $\omega _{d}$. The Hamiltonian for the
described system is given by ($\hbar =1$):
\begin{eqnarray}
H &=&\omega _{1}\hat{b}_{1}^{\dagger }\hat{b}_{1}+\omega _{2}\hat{b}%
_{2}^{\dagger }\hat{b}_{2}+J\left( \hat{b}_{1}^{\dagger }+\hat{b}_{1}\right)
\left( \hat{b}_{2}+\hat{b}_{2}^{\dagger }\right)  \notag \\
&+&F\left( \hat{b}_{1}^{\dagger }e^{i\omega _{d}t}+\hat{b}_{1}e^{-i\omega
_{d}t}\right) +U\left( \hat{b}_{2}^{\dagger }+\hat{b}_{2}\right) ^{4},
\label{H0}
\end{eqnarray}%
where $\hat{b}_{j}$ $(\hat{b}_{j}^{\dagger })$ is the annihilation
(creation) operator for the phonon mode of the $j$-th mechanical resonator
with resonance frequency $\omega _{j}$ and damp rate $\gamma _{j}$ $\left(
j=1,2\right) $. $J$ is the coupling strength between two mechanical
resonators, and $U$ is the nonlinearity of the mechanical resonator $2$. For
simplicity without loss of physics, in the following, we assume that the
mechanical resonators share the same frequency and decay rate, i.e., $\omega
_{1}=\omega _{2}=\omega _{0}$ and $\gamma _{1}=\gamma _{2}=\gamma $.
Further, we assume that the coupling strength $J$ and the mechanical
nonlinearity $U$ are much smaller than the resonance frequency $\omega _{0}$.
Under these assumptions, we can neglect the anti-rotating wave terms and
rewrite the Hamiltonian as
\begin{equation}
H=\Delta \hat{b}_{1}^{\dagger }\hat{b}_{1}+\Delta \hat{b}_{2}^{\dagger }\hat{%
b}_{2}+J\left( \hat{b}_{1}^{\dagger }\hat{b}_{2}+\hat{b}_{2}^{\dagger }\hat{b%
}_{1}\right) +F\left( \hat{b}_{1}^{\dagger }+\hat{b}_{1}\right) +U\hat{b}%
_{2}^{2\dagger }\hat{b}_{2}^{2},  \label{H}
\end{equation}%
where $\Delta =\omega _{0}-\omega _{d}$ is the detuning of the mechanical
resonance frequency from the driving frequency. The Hamiltonian (\ref{H})
describes a model of driven-dissipative coupled nonlinear mechanical
resonators, which is mathematically similar to the optical counterpart in Ref. \cite%
{UPB3}. This Hamiltonian is the starting point of our calculation.

Unlike the situation of photon blockades, the environmental temperature usually
has a significant influence on typical phonon statistics and correlation
owing to the low energy of individual phonons. With the inclusion of the
temperature factor, here, we use the Liouville-Von Neumann master equation
for the density matrix
\begin{eqnarray}
\frac{d\hat{\rho}}{dt} &=&\hat{L}\hat{\rho}  \label{L} \\
&=&-i\left[ H,\hat{\rho}\right] +\sum\limits_{n=1,2}\frac{\gamma }{2}\left[
\left( n_{th}+1\right) D\left[ \hat{b}_{n}\right] \hat{\rho}+n_{th}D\left[
\hat{b}_{n}^{\dagger }\right] \hat{\rho}\right]  \notag
\end{eqnarray}%
with the Lindblad operator $D\left[ \hat{A}\right] \hat{\rho}=2\hat{A}\hat{\rho}%
\hat{A}^{\dagger }-\hat{A}^{\dagger }\hat{A}\hat{\rho}-\hat{\rho}\hat{A}%
^{\dagger }\hat{A}$. Here, $n_{th}=\left( \exp \left( T_{0}/T\right)
-1\right) ^{-1}$ is the average phonon number of the mechanical resonators
at the temperature $T$ (We assume that the temperatures of two
nanomechanical beams are the same), and $T_{0}=\hbar \omega /K_{B}$ is the
characteristic temperature of the system with the Boltzmann constant $K_{B}$. Here, we have neglected the pure dephasing of the resonators because the dephasing rates are usually much smaller than other decay rates \cite{dephasing}.

In this case, the statistic properties of phonons for mechanical beam 1
are described by the equal-time second-order correlation function:
\begin{equation}
g^{\left( 2\right) }\left( 0\right) =\frac{\text{Tr}\left( \hat{b}%
_{1}^{\dagger }\hat{b}_{1}^{\dagger }\hat{b}_{1}\hat{b}_{1}\hat{\rho}%
_{ss}\right) }{\text{Tr}\left( \hat{b}_{1}^{\dagger }\hat{b}_{1}\hat{\rho}%
_{ss}\right) ^{2}},  \label{g2m}
\end{equation}%
where $\hat{\rho}_{ss}$ is the steady-state density matrix by setting $d\hat{%
\rho}/dt=0$ in Eq. (\ref{L}). In the calculation, we write $\hat{\rho}$ as
the density matrix $\hat{\rho}=\sum_{m,n=0}^{N}\rho _{mn,m^{\prime
}n^{\prime }}\left\vert mn\right\rangle \left\langle m^{\prime }n^{\prime
}\right\vert $ on the basis of phonon number states $\left\vert
mn\right\rangle $, where $m$ denotes the phonon number in mechanical mode 1
and $n$ denotes the phonon number in mechanical mode 2. To find the
steady-state density matrix, we need to find the eigen matrix $\hat{\rho}%
_{ss}$ of superoperator $\hat{L}$ when it has an eigenvalue of $0$, i.e., $%
\hat{L}\hat{\rho}=\lambda \hat{\rho}$ $\left( \lambda =0\right) $. Such an
eigenvalue problem can be numerically solved using the method in Ref. \cite%
{NumerMethod}. In the numerical calculation, we set the phonon number $%
n=10,m=10$, which is sufficiently large to ensure the convergence of the
results in this work.

\section{Results for Phonon Antibunching}

\subsection{Zero temperature case}

Figure 2 shows the numerical result of the equal-time second-order correlation
function as a function of the nonlinearity $U$ and mechanical coupling $J$
at zero temperature. We observe that the phonon antibunching effect occurs
in the parameter regime where the product $JU$ is relatively small.
Importantly, as opposed to the conventional phonon blockade effect, in this
coupled-mechanical oscillator system, phonons exhibit antibunching even when
the nonlinearity is negligible ($U\ll \gamma $), although this is at the cost
of a large $J$. In addition, in the regime of large $U$ and $J$, $g^{(2)}(0)$
becomes extremely large, which means that phonons are superbunched.

To find the optimal condition for phonon antibunching, we analytically solve
the steady-state master equation for the density matrix. Under the weak
driving condition $F\ll \gamma $, the average mechanical excitations would
be much lower than 1, and the Hilbert space of the total system can be
truncated to $m+n=2$. In this case, we denote $\left\vert 0,0\right\rangle
\rightarrow \left\vert 1\right\rangle $, $\left\vert 0,1\right\rangle
\rightarrow \left\vert 2\right\rangle $, $\left\vert 0,2\right\rangle
\rightarrow \left\vert 3\right\rangle $, $\left\vert 1,0\right\rangle
\rightarrow \left\vert 4\right\rangle $, $\left\vert 1,1\right\rangle
\rightarrow \left\vert 5\right\rangle $, and $\left\vert 2,0\right\rangle
\rightarrow \left\vert 6\right\rangle $. Hence, the density matrix operator $%
\hat{\rho}$ can be written as $\hat{\rho}=\sum_{m,n=1}^{6}\rho
_{mn}\left\vert m\right\rangle \left\langle n\right\vert $. In addition,
weak mechanical excitations would result in $\rho _{00}\simeq 1\gg \rho
_{01},\rho _{10}\gg \rho _{02},\rho _{20},\rho _{11}$. Then, the equal-time
second-order correlation function can be approximately expressed as:
\begin{equation}
g^{\left( 2\right) }\left( 0\right) \simeq \frac{2\rho _{66}}{\rho _{44}^{2}}%
.  \label{g2}
\end{equation}%
Obviously, if $\rho _{66}=0$, then $g^{\left( 2\right) }\left( 0\right) =0$,
which implies a low probability of having two phonons in the first
mechanical mode. After performing tedious calculations using the perturbation theory, we
obtained the expression for the matrix element $\rho _{66}$ in the zero-temperature case ($n_{th}=0$)
\begin{equation}
\rho _{66}=\frac{F^{4}\left\vert U\left( J^{2}+2\tilde{\Delta}^{2}\right) +2%
\tilde{\Delta}^{3}\right\vert ^{2}}{2J^{8}\left\vert U+2\tilde{\Delta}%
\right\vert ^{2}}.  \label{r66}
\end{equation}%
where $\tilde{\Delta}=\Delta -i\gamma /2$. In the calculation, we further assume a large mechanical coupling $J$; otherwise, the
expression for matrix elements would be more complicated. Then, we can find the
following condition for perfect antibunching
\begin{eqnarray}
2UJ^{2}+4\left( \Delta +U\right) \Delta ^{2}-\left( 3\Delta +U\right) \gamma
^{2} &=&0  \label{D0} \\
8U\Delta +12\Delta ^{2}-\allowbreak \gamma ^{2} &=&0  \label{D1}
\end{eqnarray}%
which is mathematically the same as the results in the photon blockade case
\cite{UPB3}. We plot the optimal condition in Fig. 2 (white dashed curve),
which is in good agreement with the numerical result.

In the situation where the mechanical coupling $J$ is much greater than the
mechanical damping $\gamma $, the optimal parameters required to achieve $\rho
_{66}=0 $ can be approximated as
\begin{eqnarray}
\Delta _{optimal} &\simeq &\frac{\gamma }{2\sqrt{3}}  \label{U} \\
U_{optimal} &\simeq &\frac{2\gamma ^{3}}{3\sqrt{3}J^{2}}
\end{eqnarray}%
Obviously, in this case, the nonlinearity that is required to observe an optimal
blockade can be made extremely small ($U_{optimal}/\gamma =2\gamma
^{2}\left/ \left( 3\sqrt{3}J^{2}\right) \right. \ll 1$), contrary to the
condition of conventional phonon blockade ($U\left/ \gamma \right. \gg 1$).
\begin{figure}[tbp]
\onefigure[width=3.2in]{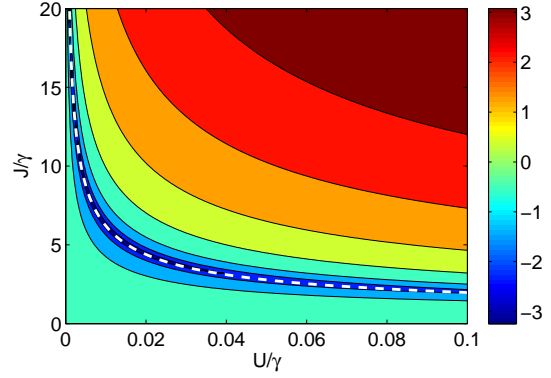}
\caption{{(Color online) Equal-time second-order correlation function $
\mbox{log}_{10}(g^{(2)}(0))$ as a function of the mechanical coupling $J$
and the nonlinear coefficient $U$ at zero temperature. The white dashed
curve is calculated from the analytical optimal conditions Eqs. (\protect\ref%
{D0}) and (\protect\ref{D1}). The parameters are: $\protect\gamma =1 $, $%
\Delta =0.29\protect\gamma $, and $F=0.001\protect\gamma $.}}
\end{figure}

We then study the phonon number probability distribution $%
P_{m}=\sum_{n=0}^{\infty }\rho _{mn,mn}$ of mechanical mode 1 when the
phonon blockade occurs. For comparison, we also show the probability
distribution of a coherent state with the same mean phonon number $%
\left\vert \alpha \right\vert ^{2}=\sum_{m=0}^{\infty }P\left( m\right) $,
which obeys the Poisson distribution $P\left( n\right) =\left\vert \alpha
\right\vert ^{2n}e^{-\left\vert \alpha \right\vert ^{2}}/n!$. As shown in
Fig. 3, the largest probability is occupied by the vacuum state. The
one-phonon state is the next most probable. The two-phonon state is significantly
less probable than the single-phonon state. Compared to the coherent state
(the same mean phonon number with the blockade case), the probability for
the two-phonon state is small, which indicates that the state of mechanical mode 1 has a nonclassical distribution. Recently, similar results
have also been observed in optical counterparts\cite{UPB0}.

\begin{figure}[tbp]
\onefigure[width=3.2in]{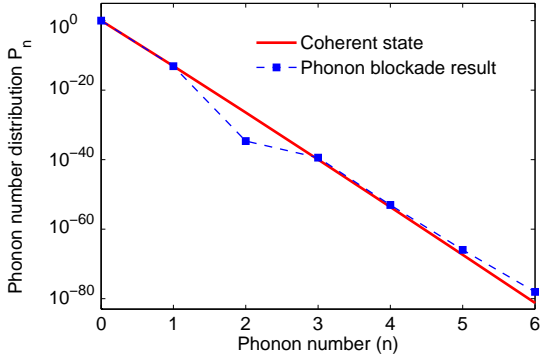}
\caption{{(Color online) Population distribution of number state of phonon
mode 1. The other parameters are: $\protect\gamma =1$, $U=0.00387\protect%
\gamma$, $J = 10\protect\gamma$, $\Delta =0.2874\protect\gamma$, and $F =
0.00005\protect\gamma$.}}
\label{popu}
\end{figure}

\subsection{Finite temperature case}

In the previous subsection, we studied the phonon blockade effect without
including the effect of environmental temperature. However, cooling a
mechanical resonator down to zero temperature is not easy achieved experimentally.
Therefore, it is important to study the phonon blockade at finite
temperature. First, in Fig. 4, we numerically plot $g^{(2)}(0)$ as a function of $J$ and
$U$ at $T=0.04T_{0}$. It can be seen that the phonon antibunching
effect still exists in the weak nonlinearity $U$, but small mechanical
coupling $J$ regime (enclosed by the white solid line in Fig. 4). It is also observed that there is strong phonon antibunching in a specific regime,
which is similar to that in the zero-temperature case. However, in the regime
of weak nonlinearity $U$ and large $J$, $g^{(2)}(0)$ is greater than 1,
which is contrary to the zero-temperature case (see Fig. 2). This behavior
indicates that thermal noise destroys the phonon antibunching effect in
the regime of weak nonlinearity $U$ and large $J$.

\begin{figure}[tbp]
\onefigure[width=3.2in]{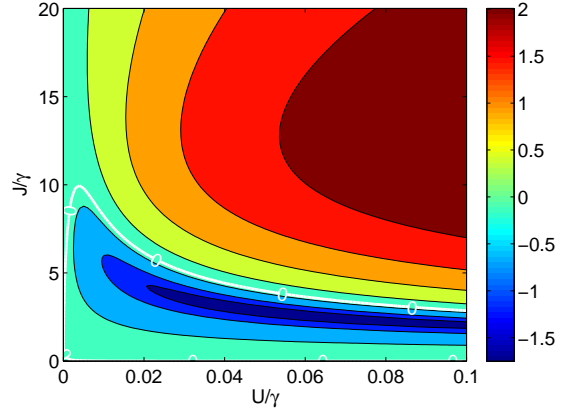}
\caption{{(Color online) Equal-time second-order correlation function $
\mbox{log}_{10}(g^{(2)}(0))$ as a function of mechanical coupling $J$ and
nonlinearity $U$ at temperature $T=0.04T_0$. The solid white line indicates
the contour line with $g^{(2)}(0)=1$. The other parameters are: $\protect%
\gamma =1$, $\Delta =0.29\protect\gamma$ and $F=0.001\protect\gamma$.}}
\end{figure}

To study the effect of the environmental temperature on the phonon blockade,
we present the analytical expressions of the density matrix elements $\rho
_{66}$ and $\rho _{44}$ in the strong coupling regime (Using the same
procedure as in the zero-temperature case):
\begin{eqnarray}
\rho _{44} &=&\frac{F^{2}\left\vert \tilde{\Delta}\right\vert ^{2}}{%
\left\vert J^{2}-\tilde{\Delta}^{2}\right\vert ^{2}}+n_{th}  \label{44} \\
\rho _{66} &=&\frac{F^{4}\left\vert U\left( J^{2}+2\tilde{\Delta}^{2}\right)
+2\tilde{\Delta}^{3}\right\vert ^{2}}{2J^{8}\left\vert U+2\tilde{\Delta}%
\right\vert ^{2}}+\frac{2F^{2}\left\vert \tilde{\Delta}\right\vert ^{2}}{%
J^{4}}n_{th}+n_{th}^{2}  \label{66}
\end{eqnarray}%
From Eqs. (\ref{44}-\ref{66}), both $\rho _{44}$ and $\rho _{66}$ contain
two parts: one part determined by the temperature or the thermal phonon
number, and the other part from the system itself (quantum interference).
These two parts incoherently contribute to the phonon correlation. From Eq. (%
\ref{66}), it can be noted that $\rho _{66}>0$, which implies that the perfect
antibunching condition no longer exists at finite temperature. However, from
Eq. (\ref{66}) we can see that for a fixed temperature $T$, when the first
term vanishes, the equal-time second-order correlation will take the
minimal value. Therefore, the optimal condition for the strong phonon
antibunching at finite temperature is the same as that at zero
temperature. When the environmental temperature is small, the region for $%
0<g^{(2)}(0)<1$ still exists because the quantum interference partially
contributing to phonon correlation leads to an imperfect antibunching
effect. When the mechanical coupling $J$ is small, the contribution from
quantum interference is enhanced, leading to a stronger phonon antibunching,
and vice versa. These conclusions are well confirmed by the numerical
results in Fig. 4.

\begin{figure}[tbp]
\onefigure[width=3.2in]{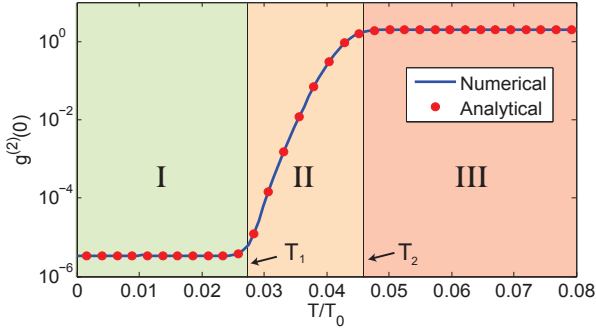}\label{g_T}
\caption{{(Color online) Equal-time second-order correlation function $%
g^{(2)}(0)$ as a function of environmental temperature $T$. I, II, and III
represent the quantum regime, crossover regime, and thermal regime,
respectively. The numerical result is obtained based on the steady-state master
equation (\protect\ref{L}), and the analytical result is calculated based on
Eqs. (\protect\ref{g2}, \protect\ref{44}, \protect\ref{66}). The parameters
that are used in this figure are: $\protect\gamma =1$, $U=0.00096\protect\gamma $%
, $J=20\protect\gamma $, $\Delta =0.2885\protect\gamma $, and $F=0.01\protect%
\gamma $.}}
\end{figure}

Figure 5 shows $g^{\left( 2\right) }\left( 0\right) $ as a function of the
environmental temperature. Based on different behaviors of the second-order
correlation function, the whole regime can be divided into three parts by
two boundary temperatures $0.028T_{0}$ and $0.046T_{0}$. In the first region
(region I), $g^{2}(0)$ is a constant and is much smaller than $1$, which means that the
phononic mode has a strong antibunching effect. Therefore, we called it the
quantum regime. Interestingly, the quantum correlation is hardly affected by
the thermal noise in the quantum regime, which is very helpful to experimentally
observe the strong phonon antibunching at low finite temperature. In the
second region (region II), $g^{\left( 2\right) }\left( 0\right) $
experiences a sharp increase from $3\times 10^{-6}$ to $2$ with increasing
temperature. Obviously, in this regime, the quantum correlation of the phononic mode
undergoes a transition from strong antibunching to bunching.
We call it the crossover regime. In the third region (region III), $%
g^{\left( 2\right) }\left( 0\right) $ approaches another constant $2$, which
means that the phononic mode has the statistic property of a thermal field.
This regime is called the thermal regime.

These phenomena described in last paragraph can be explained from Eq. (\ref%
{g2}) and (\ref{66}) as follows. In the quantum regime, the first term is much
larger than other terms in Eq. (\ref{66}), i.e., quantum
interference dominates, and therefore, $g^{2}(0)$ mainly shows the pure
quantum correlation. In the thermal regime, the third term is much larger
than other terms in Eq. (\ref{66}), i.e., pure thermal noise dominates,
which results in $g^{\left( 2\right) }\left( 0\right) \approx 2$. In the
crossover regime, the gradually enhanced thermal noise gradually suppresses
and eventually destroys the quantum correlation; hence, $g^{2}(0)$ monotonously
increases from an almost vanishing value to $2$. Two critical temperatures
can be determined as follows: when the first term is equal to the second
term and much greater than the third term in Eq. (\ref{66}), the
corresponding temperature is the first boundary value. Then, we find $%
T_{1}=T_{0}/\ln \left( 1+4J^{4}\left\vert U\tilde{\Delta}+2\tilde{\Delta}%
^{2}\right\vert ^{2}/F^{2}\left\vert U\left( J^{2}+2\tilde{\Delta}%
^{2}\right) +2\tilde{\Delta}^{3}\right\vert ^{2}\right) $; when the third
term is equal to the second term and much greater than the first term in Eq.
(\ref{66}), we find another boundary temperature $T_{2}=T_{0}/\ln \left(
1+J^{4}/2\left\vert F\tilde{\Delta}\right\vert ^{2}\right) $.

\begin{figure}[tbp]
\onefigure[width=3.2in]{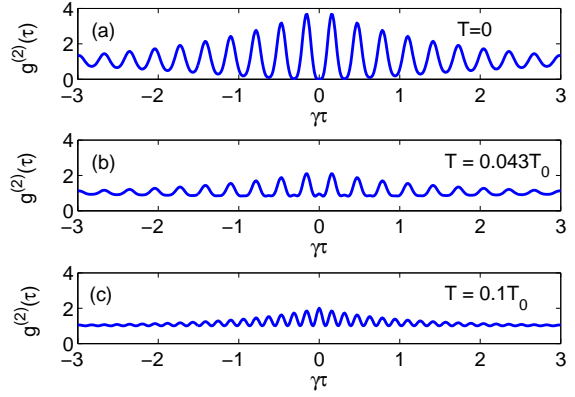}
\caption{{(Color online) (a-c) Second-order correlation function $g^{(2)}(%
\protect\tau)$ is plotted as a function of time delay $\protect\tau $ at
different temperatures. The parameters used in this figure are: $\protect%
\gamma =1$, $U=0.00096\protect\gamma $, $J=20\protect\gamma $, $\Delta =0.288%
\protect\gamma $, and $F=0.01\protect\gamma $.}}
\end{figure}

We also studied the second-order correlation function versus the time delay $%
g^{(2)}(\tau )$ at three different environment temperatures, as shown in
Fig. 6. Owing to the probability oscillation between the phonon state $%
\left\vert 0,1\right\rangle $ and $\left\vert 1,0\right\rangle $, the second-order correlation functions oscillate with period $2\pi /J$. In the zero-temperature case (Fig. 6(a)), the value of $g^{\left( 2\right) }\left(
0\right) $ becomes negligible. For $T=0.043T_{0}$ (Fig. 6(b)), the
amplitude of the oscillation of $g^{\left( 2\right) }\left( \tau \right) $
decreases; meanwhile, the value of $g^{\left( 2\right) }\left( 0\right) $
increases to unitary. When the temperature is further increased to $%
T=0.1T_{0}$ (Fig. 6(c)), the value of $g^{\left( 2\right) }\left( 0\right) $
is $\simeq 2$, indicating that the phonon field in the mechanical resonator $1$
is a thermal field. Again, the time evolution of the second-order correlation
function shows that a large thermal noise will suppress and even destroy the
phonon antibunching. From Fig. 6, one can see that strong phonon
antibunching is demonstrated in a range about $\pi /J$ around the
zero-delay, which gives a limited requirement for the temporal resolution of
a detector.

\begin{figure}[tbp]
\onefigure[width=3.2in]{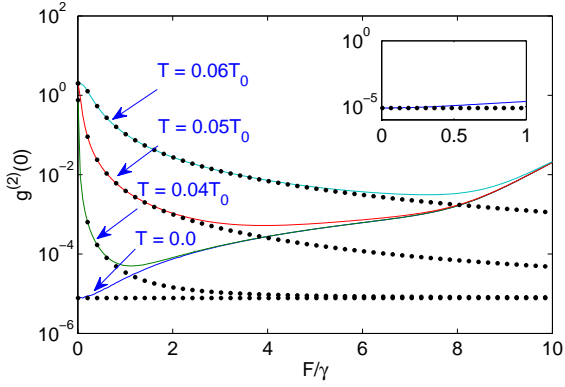}
\caption{{(Color online) Equal-time second-order correlation function $%
g^{(2)}(0)$ versus the driving strength F at different temperatures. The dotted line
represents analytical results, and the solid line represents the numerical results.
The inset is the zero-temperature case. The other parameters are: $\protect%
\gamma=1$, $U=0.00096\protect\gamma$, $J=20\protect\gamma$, $\Delta=0.288%
\protect\gamma$.}}
\end{figure}

We then investigate the influence of the driving force on the phonon
statistics. In Fig. 7, we plot $g^{(2)}(0)$ of the phononic mode in
mechanical resonator 1 as a function of the driving force at zero and finite
temperatures. First, we focus on the zero-temperature case.
From Eqs. (\ref{g2}), (\ref{44}), and (\ref{66}) we can obtain the equal-time
second-order correlation function at zero temperature:
\begin{equation}
g^{\left( 2\right) }\left( 0\right) =\left\vert \frac{\left( U\left( J^{2}+2%
\tilde{\Delta}^{2}\right) +2\tilde{\Delta}^{3}\right) \left\vert J^{2}-%
\tilde{\Delta}^{2}\right\vert ^{2}}{\left( U+2\tilde{\Delta}\right)
\left\vert \tilde{\Delta}\right\vert ^{2}J^{4}}\right\vert ^{2}.  \label{g20}
\end{equation}%
It should be noted that in the weak driving regime, $g^{(2)}(0)$ is independent of $F$,
and is consistent with the numerical result shown in Fig. 7. However, the
analytical result clearly deviates from the numerical result in the strong
driving regime. It is shown that the analytical result is valid only in the
weak driving regime. The reason is that a strong driving enhances the
population in the high phonon number states, and therefore decreases the
antibunching effect.

Next, we discuss the influence of the driving amplitude $F$ on $g^{(2)}(0)$
in the finite-temperature case. As opposed to the zero-temperature case, it
can be seen that all second-order correlation curves at finite temperature
first decrease and then increase as the driving force is enhanced. Indeed,
a similar phonon correlation behavior was numerically observed without
physical interpretation in Ref. \cite{Phonon2}. In the weak driving limit,
the analytical model well reproduces the numerical calculation. However,
because the analytical model is based on the assumption that the total phonon number is
small ($m+n=2$), it fails to reproduce the increasing $g^{(2)}(0)$ at higher
driving. These observations can be understood from Eqs. (\ref{g2}), (\ref{44}%
), and (\ref{66}). When $F$ is very small, the thermal noise term dominates
the matrix elements $\rho _{44}$ and $\rho _{66}$, and hence, the phonon state
is approximately a thermal state ($g^{(2)}(0)\approx 2$). With the increase
in driving strength, the contributions from the quantum-interference terms
in Eqs. (\ref{44}) and (\ref{66}) become important, leading to a decrease in
$g^{(2)}(0)$. When the driving is sufficiently strong, the three and higher phonon
states in resonator $1$ are populated with increased probability owing to
the strong coupling between resonator $1$ and resonator $2$. Hence, the
second-order correlation function will increase again, resulting in a
degradation of phonon antibunching. From the above discussions, it is clear
that a very weak force is required to achieve a stronger antibunching effect
for the zero-temperature case, while a suitably strong driving strength is
helpful to preserve the stronger phonon antibunching effect when thermal
noise exists.

Now, we discuss the experimental feasibility to observe the phonon blockade
effect in the presence of unavoidable thermal noise. We assume that the two
resonators in our model are silicon-based MEMS/NEMS resonators, which are
doubly clamped rectangular cross-section beams with width $d=5$ nm and length
$L=100$ nm. According to Ref. \cite{mems,coupling1}, we find that the
fundamental frequency is $\omega _{0}/2\pi =d/L^{2}\sqrt{E/\rho }=4.3$ GHz
and the nonlinear strength $U=\hbar \left( 1.6\rho d^{4}L\right) ^{-1}=430$
Hz. The decay rate of the resonator is dependent on many factors, such as the
geometrical configuration, anchor loss, environmental pressure  and temperature,
input-output loss, and thermoelastic damping. With a suitable design and fabrication, the
total decay can be controlled. For simplicity, we take a typical value of $%
\gamma =10$ MHz, which corresponds to a quality factor $Q\sim 4300$. If we
assume the coupling between two mechanical resonators comes from an
electrostatic force, the coupling strength $J$ can be freely adjusted by
changing the externally applied voltage. With these feasible parameters, we
can then estimate the characteristic temperature $T_{0}=\hbar \omega
_{0}/K_{B}\sim 196$ mK. According to the previous discussion, for
an environmental temperature below $5.3$ mK, the phonon blockade is almost
unaffected by the thermal noise. Such a temperature is attainable using
current laboratory technology. To measure the time-delayed second-order
correlation function $g^{(2)}(\tau )$, we can adapt the same method as in Ref. \cite{HBT}. In this
reference \cite{HBT}, the phonon correlation can be transferred to the photon correlation,
which can be measured by a typical photonic Hanbury Brown and Twiss (HBT) set-up. Here,
we take coupling strength $J=20\gamma \sim $ $200$ MHz, corresponding to a
window width of phonon antibunching $\pi /J\sim 10$ ns. Apparently, this time scale is much greater
than a typical temporal resolution of the detector (usually $<100$ ps)\cite{time_r}.

\section{Conclusion}

\label{conl} We studied the phonon antibunching effect in a coupled
mechanical resonator system with weak mechanical nonlinearity both at zero
and finite temperatures. In the weak driving limit, analytical results were derived for the phonon antibunching in the strong coupling regime, and
the temperature effect on the phonon blockade was investigated based on the
stationary master equation. The results show that there exists a perfect
phonon antibunching when the system parameters take the optimal conditional
value at zero temperature. When thermal noise exists, the perfect
antibunching condition is destroyed, leading to a suppressed phonon
antibunching effect in the regime of small mechanical coupling, or leading to
a destroyed phonon antibunching in the regime of large mechanical coupling.
Furthermore, we find that the thermal noise has a different effect on the phonon
correlation in the quantum regime, crossover regime, and thermal regime.
Temperature variations have a negligible influence on the quantum
correlation of the phonon mode when the environmental temperature is present in the
quantum regime. Finally, the effect of the driving force on the
phonon blockade is also discussed. It is found that a suitably strong
driving, rather than a very weak driving, would help to preserve an optimal
phonon antibunching effect at finite temperature.

\acknowledgments
This work is supported by the National Natural Science Foundation of China
under Grants No. 11364021 and No. 11664014, Natural Science Foundation of
Jiangxi Province under Grant 20161BAB201023. W.P. Bowen acknowledges
support from the Australian Research Council through grants CE110001013 and
FT140100650.

\end{document}